\documentclass[11pt]{article}
\pdfoutput=1

\usepackage{amsmath,amsfonts,amsbsy,amssymb,array,accents,dsfont} 
\usepackage{enumerate,latexsym,graphicx,mathrsfs,verbatim,psfrag,xspace} 
\usepackage{bm} 
\usepackage[normalem]{ulem}
\usepackage{booktabs} 
\usepackage[usenames]{color}

\usepackage[english]{babel} 
\usepackage{fancybox} 
\usepackage{slashed}  

\usepackage[utf8]{inputenc}

\usepackage[nosort]{cite}
\usepackage{chngpage} 

\usepackage[nosort]{cite}   
\usepackage{mciteplus}

\usepackage[colorlinks=true,      linkcolor=blue,      urlcolor=blue,      filecolor=blue,      citecolor=blue,
      pdfstartview=FitH,     pdfpagemode=UseNone,      bookmarksopen=true]{hyperref}  
      
\usepackage[all]{hypcap}     

\newcommand{\captionfonts}{\small}

\makeatletter  
\long\def\@makecaption#1#2{%
  \vskip\abovecaptionskip
  \sbox\@tempboxa{{\captionfonts #1: #2}}%
 \ifdim \wd\@tempboxa >\hsize
    {\captionfonts #1: #2\par}
  \else
    \hbox to\hsize{\hfil\box\@tempboxa\hfil}%
  \fi
  \vskip\belowcaptionskip}
\makeatother   

\begin{document}

\numberwithin{equation}{section}


\mathchardef\mhyphen="2D


\newcommand{\be}{\begin{equation}}
\newcommand{\ee}{\end{equation}}
\newcommand{\bea}{\begin{eqnarray}\displaystyle}
\newcommand{\eea}{\end{eqnarray}}
\newcommand{\nnm}{\nonumber}
\newcommand{\nn}{\nonumber}
\newcommand{\newotimes}{}  
\newcommand{\utv}{|0_R^{-}\rangle^{(1)}\newotimes |0_R^{-}\rangle^{(2)}}
\newcommand{\utvb}{|\bar 0_R^{-}\rangle^{(1)}\newotimes |\bar 0_R^{-}\rangle^{(2)}}

\def\eq#1{(\ref{#1})}
\newcommand{\secn}[1]{Section~\ref{#1}}

\newcommand{\tbl}[1]{Table~\ref{#1}}
\newcommand{\fig}{Fig.~\ref}

\def\beq{\begin{equation}}
\def\eeq{\end{equation}}
\def\beqa{\begin{eqnarray}}
\def\eeqa{\end{eqnarray}}
\def\bet{\begin{tabular}}
\def\eet{\end{tabular}}
\def\bs{\begin{split}}
\def\es{\end{split}}
\def\sqi{{1\over \sqrt{2}}}

\definecolor{cardinal}{rgb}{0.6,0,0}
\definecolor{darkgreen}{rgb}{0,0.4,0}
\definecolor{purple}{rgb}{0.5, 0, 0.5}
\definecolor{golden}{rgb}{0.92, 0.7, 0}
\definecolor{midnight}{rgb}{0, 0, 0.5}
\definecolor{darkblue}{rgb}{0, 0, 0.7}
\newcommand{\Red}{\color{red}}
\newcommand{\Blue}{\color{blue}}
\newcommand{\Purple}{\color{purple}}
\newcommand{\DarkGreen}{\color{darkgreen}}
\newcommand{\Cardinal}{\color{cardinal}}
\newcommand{\Golden}{\color{golden}}
\newcommand{\Midnight}{\color{midnight}}
\newcommand{\DarkBlue}{\color{darkblue}}
\def\DG{\DarkGreen}
\def\CR{\Cardinal}
\def\DB{\DarkBlue}


\def\DT#1{{\bf \Red DT:} {\DarkGreen #1}}
\def\SM#1{{\bf \Red SDM:} {\DB #1}}
\def\SDM#1{{\bf \Red SDM:} {\DB #1}}


\def\a{\alpha}  \def\b{\beta}   \def\c{\chi}    
\def\g{\gamma}  \def\G{\Gamma}  \def\e{\epsilon}  
\def\vep{\varepsilon}   \def\tvep{\widetilde{\varepsilon}}
\def\f{\phi}    \def\F{\Phi}  \def\fb{{\ov \phi}}
\def\vf{\varphi}  \def\m{\mu}  \def\mub{\ov \mu}
\def\n{\nu}  \def\nub{\ov \nu}  \def\o{\omega}
\def\O{\Omega}  \def\r{\rho}  \def\k{\kappa}
\def\kab{\ov \kappa}  \def\s{\sigma}
\def\t{\tau}  \def\th{\theta}  \def\sb{\ov\sigma}  \def\S{\Sigma}
\def\l{\lambda}  \def\L{\Lambda}  \def\p{\psi}

\newcommand{\gt}{\tilde{\gamma}}


\def\cA{{\cal A}} \def\cB{{\cal B}} \def\cC{{\cal C}}
\def\cD{{\cal D}} \def\cE{{\cal E}} \def\cF{{\cal F}}
\def\cG{{\cal G}} \def\cH{{\cal H}} \def\cI{{\cal I}}
\def\cJ{{\cal J}} \def\cK{{\cal K}} \def\cL{{\cal L}}
\def\cM{{\cal M}} \def\cN{{\cal N}} \def\cO{{\cal O}}
\def\cP{{\cal P}} \def\cQ{{\cal Q}} \def\cR{{\cal R}}
\def\cS{{\cal S}} \def\cT{{\cal T}} \def\cU{{\cal U}}
\def\cV{{\cal V}} \def\cW{{\cal W}} \def\cX{{\cal X}}
\def\cY{{\cal Y}} \def\cZ{{\cal Z}}

\def\mC{\mathbb{C}} \def\mP{\mathbb{P}}  
\def\mR{\mathbb{R}} \def\mZ{\mathbb{Z}} 
\def\mT{\mathbb{T}} \def\mN{\mathbb{N}}
\def\mH{\mathbb{H}} \def\mX{\mathbb{X}}

\def\CP{\mathbb{CP}}
\def\RP{\mathbb{RP}}
\def\Z{\mathbb{Z}}
\def\N{\mathbb{N}}
\def\H{\mathbb{H}}

\newcommand{\rmd}{\mathrm{d}}
\newcommand{\rmx}{\mathrm{x}}

\def\tA{ {\widetilde A} } 

\def\one{{\hbox{\kern+.5mm 1\kern-.8mm l}}}
\def\zero{{\hbox{0\kern-1.5mm 0}}}


\newcommand{\bra}[1]{{\langle {#1} |\,}}
\newcommand{\ket}[1]{{\,| {#1} \rangle}}
\newcommand{\braket}[2]{\ensuremath{\langle #1 | #2 \rangle}}
\newcommand{\Braket}[2]{\ensuremath{\langle\, #1 \,|\, #2 \,\rangle}}
\newcommand{\norm}[1]{\ensuremath{\left\| #1 \right\|}}
\def\corr#1{\left\langle \, #1 \, \right\rangle}
\def\vac{|0\rangle}


\def\d{ \partial } 
\def\zb{{\bar z}}

\newcommand{\sq}{\square}
\newcommand{\IP}[2]{\ensuremath{\langle #1 , #2 \rangle}}    

\newcommand{\floor}[1]{\left\lfloor #1 \right\rfloor}
\newcommand{\ceil}[1]{\left\lceil #1 \right\rceil}

\newcommand{\dbyd}[1]{\ensuremath{ \frac{\d}{\d {#1}}}}
\newcommand{\ddbyd}[1]{\ensuremath{ \frac{\d^2}{\d {#1}^2}}}

\newcommand{\Zd}{\ensuremath{ Z^{\dagger}}}
\newcommand{\Xd}{\ensuremath{ X^{\dagger}}}
\newcommand{\Ad}{\ensuremath{ A^{\dagger}}}
\newcommand{\Bd}{\ensuremath{ B^{\dagger}}}
\newcommand{\Ud}{\ensuremath{ U^{\dagger}}}
\newcommand{\Td}{\ensuremath{ T^{\dagger}}}

\newcommand{\T}[3]{\ensuremath{ #1{}^{#2}_{\phantom{#2} \! #3}}}		

\newcommand{\tr}{\operatorname{tr}}
\newcommand{\sech}{\operatorname{sech}}
\newcommand{\Spin}{\operatorname{Spin}}
\newcommand{\Sym}{\operatorname{Sym}}
\newcommand{\Com}{\operatorname{Com}}
\def\adj{\textrm{adj}}
\def\id{\textrm{id}}

\def\ha{\frac{1}{2}}
\def\tha{\tfrac{1}{2}}
\def\wt{\widetilde}
\def\ra{\rangle}
\def\la{\langle}

\def\pb{\ov\psi}
\def\pt{\widetilde{\psi}}
\def\at{\widetilde{\a}}
\def\cb{\ov\chi}
\def\d{\partial}
\def\db{\bar\partial}
\def\delb{\bar\partial}
\def\dbar{\ov\partial}
\def\dag{\dagger}
\def\dalpha{{\dot\alpha}}
\def\dbeta{{\dot\beta}}
\def\dgamma{{\dot\gamma}}
\def\ddelta{{\dot\delta}}
\def\ad{{\dot\alpha}}
\def\bd{{\dot\beta}}
\def\dg{{\dot\gamma}}
\def\dd{{\dot\delta}}
\def\th{\theta}
\def\Th{\Theta}
\def\eb{{\ov \epsilon}}
\def\gb{{\ov \gamma}}
\def\wb{{\ov w}}
\def\Wb{{\ov W}}
\def\D{\Delta}
\def\DD{\Delta^\dag}
\def\Db{\ov D}

\def\ov{\overline}
\def\Slash{\, / \! \! \! \!}
\def\dslash{\partial\!\!\!/} 
\def\Dslash{D\!\!\!\!/\,\,}
\def\fslash#1{\slash\!\!\!#1}
\def\Fslash#1{\slash\!\!\!\!#1}

\def\del{\partial}
\def\delb{\bar\partial}
\newcommand{\ex}[1]{{\rm e}^{#1}} 
\def\ii{{i}}

\newcommand{\vs}[1]{\vspace{#1 mm}}

\newcommand{\ve}{{\vec{\e}}}
\newcommand{\shalf}{\frac{1}{2}}

\newcommand{\lb}{\rangle}
\newcommand{\al}{\ensuremath{\alpha'}}
\newcommand{\ap}{\ensuremath{\alpha'}}

\newcommand{\bean}{\begin{eqnarray*}}
\newcommand{\eean}{\end{eqnarray*}}
\newcommand{\ft}[2]{{\textstyle {\frac{#1}{#2}} }}

\newcommand{\hsp}{\hspace{0.5cm}}
\def\half{{\textstyle{1\over2}}}
\let\ci=\cite \let\re=\ref
\let\se=\section \let\sse=\subsection \let\ssse=\subsubsection

\newcommand{\dpb}{D$p$-brane}
\newcommand{\dpbs}{D$p$-branes}

\def\gh{{\rm gh}}
\def\sgh{{\rm sgh}}
\def\NS{{\rm NS}}
\def\R{{\rm R}}
\def\Qp{Q_{\rm P}}
\def\QP{Q_{\rm P}}

\newcommand\dott[2]{#1 \! \cdot \! #2}

\def\eo{\overline{e}}


\def\p{\partial}
\def\h{{1\over 2}}

\def\d{\partial}
\def\la{\lambda}
\def\eps{\epsilon}
\def\bb{\bigskip}
\def\tg{\widetilde\gamma}
\newcommand{\dm}{\begin{displaymath}}
\newcommand{\edm}{\end{displaymath}}
\renewcommand{\b}{\widetilde{B}}
\newcommand{\gm}{\Gamma}
\newcommand{\ac}[2]{\ensuremath{\{ #1, #2 \}}}
\renewcommand{\ell}{l}
\newcommand{\z}{\ell}
\def\bb{$\bullet$}
\def\Qbar{{\bar Q}_1}
\def\QPbar{{\bar Q}_p}

\def\q{\quad}

\def\bn{B_\circ}

\let\a=\alpha \let\b=\beta \let\g=\gamma 
\let\e=\epsilon
\let\c=\chi \let\th=\theta  \let\k=\kappa
\let\l=\lambda \let\m=\mu \let\n=\nu \let\x=\xi \let\r=\rho

\let\s=\sigma

\let\vp=\varphi \let\vep=\varepsilon
\let\w=\omega  \let\G=\Gamma \let\D=\Delta \let\Th=\Theta \let\P=\Pi \let\S=\Sigma

\def\h{{1\over 2}}

\def\r{\rightarrow}
\def\Ri{\Rightarrow}

\def\nn{\nonumber\\}
\let\bm=\bibitem
\def\Kt{{\widetilde K}}
\def\b{\vspace{3mm}}
\def\t{\tilde}
\let\p=\partial

\newcommand{\nsonep}{NS1-P\xspace}
\newcommand{\donep}{D1-P\xspace}
\newcommand{\donedfive}{D1-D5\xspace}
\newcommand{\donedfivep}{D1-D5-P\xspace}

\newcommand{\NSNS}{\ensuremath{\textsc{NSNS}}}
\newcommand{\RR}{\ensuremath{\textrm{RR}}}
\newcommand{\el}{\ensuremath{\:\!\textrm{el}}}
\renewcommand{\mag}{\ensuremath{\textrm{mag}}}

\newcommand{\BH}{\ensuremath{\textsc{BH}}}
\newcommand{\micro}{\ensuremath{\textrm{micro}}}


\begin{flushright}
\end{flushright}

\vspace{20mm}

 \begin{center}
{\LARGE The fuzzball nature of two-charge black hole microstates}
\\

\vspace{20mm}

{\bf  Samir D. Mathur${}^{1}$~ and ~ David Turton${}^{2}$}

\vspace{13mm}

${}^{1}$ Department of Physics,\\ The Ohio State University,\\ Columbus,
OH 43210, USA\\  \vskip 1mm  mathur.16@osu.edu\\

\vspace{8mm}

${}^2$Mathematical Sciences and STAG Research Centre, \\ University of Southampton,\\ Highfield,
Southampton SO17 1BJ, UK\\ \vskip 1mm  d.j.turton@soton.ac.uk\\

\end{center}

\vspace{15mm}

\begin{abstract}

\vspace{0.4 cm}

\noindent

It has been suggested by A.~Sen that the entropy of two-charge supersymmetric bound states in string theory should be accounted for by adding the entropy of source-free horizonless supergravity solutions to the entropy associated with the horizons of small black holes. This would imply that the entropy arises differently depending on the duality frame: in the D1-D5 frame one would count source-free horizonless solutions, while in the NS1-P frame one would compute the area of a horizon. This might lead to the belief that the microstates are described by fuzzball solutions in the D1-D5 duality frame but by a black hole with a horizon in the latter. We argue that this is not the case, and that the microstates are fuzzballs in all duality frames. We observe that the scaling argument used by Sen fails to account for the entropy in the D1-P and other duality frames. We also note that the traditional extremal black hole solution is not a complete string background, since finite-action paths connect the exterior near-horizon extremal throat to the region inside the horizon, including the singularity. The singularity of the traditional black hole solution does not give a valid boundary condition for a fundamental string; correcting this condition by resolving the singularity modifies the black hole to a fuzzball with no horizon. We argue that for questions of counting states, the traditional black hole solution should be understood through its Euclidean continuation as a saddle point, and that the Lorentzian states being counted are fuzzballs in all duality frames.

\end{abstract}

\thispagestyle{empty}

\newpage

\section{Introduction}
\label{intr}\setcounter{footnote}{0}

\baselineskip=15pt
\parskip=3pt

String theory has seen significant success in the exploration of the fundamental physics of black holes.
Supersymmetric (BPS) black holes carrying two conserved charges have provided a particularly fruitful laboratory.
The entropy of BPS states of the heterotic string carrying $n_w$ units of winding and $n_p$ units of momentum was first computed by Sen~\cite{Sen:1994eb,Sen:1995in}. The microscopic entropy at large  $n_wn_p$, 
\be
S_{\micro}\;\simeq\; 4\pi \sqrt{n_wn_p} \;,
\label{one}
\ee
gave the first example of a microscopic count of states which reproduces black hole entropy in string theory. The spherically symmetric BPS black hole solution with these charges is singular in supergravity. However Sen argued that higher-derivative corrections would be relevant at some fixed distance from the singularity; placing a stretched horizon there gave a gravitational entropy proportional to the area $A$ of the stretched horizon, $S_{\mathrm{grav}}\;\sim\; {A\over G}$, yielding the value
\bea
S_{\mathrm{grav}}\;=\;c \sqrt{n_wn_p} \;,
\label{two}
\eea
where $c$ is a  constant of order unity whose value depends on the precise location of the stretched horizon. Later work  found that the constant $c$ in (\ref{two}) agreed with (\ref{one}) for the heterotic string~\cite{Dabholkar:2004yr}, but for similar states in Type IIA/IIB theory the constant $c$ was found to vanish.

The states counted in (\ref{one}) certainly exist in string theory, so one can ask: what is their gravitational description when $n_w, n_p \gg 1$? This question was addressed in \cite{Lunin:2001jy} for states in Type IIB theory compactified as 
\bea
M_{9,1}\;\r \;M_{4,1}\times S^1\times T^4 \,,
\label{eq:compactification}
\eea
where $M_{d,1}$ denotes ($d$+1)-dimensional Minkowski space asymptotics. 
The entropy (\ref{one}) arises from the different ways that the momentum can be distributed among the various harmonics of the string. One can start by considering simple states, in which a few harmonics $k$ are populated with excitation numbers $n_k\gg 1$. Coherent states of this type are described by a classical vibration profile $\vec F(\vec x)$ describing the motion of the string, and the gravitational solution for the string carrying such a wave can be computed~\cite{Lunin:2001jy,Lunin:2002iz,Mathur:2005zp,Taylor:2005db,Kanitscheider:2006zf,Kanitscheider:2007wq}. These solutions, given in (\ref{eq:ns1p-sol}) below, turn out to have no horizon.  They also do not have the kind of singularity found inside a black hole. The supergravity fields instead have the natural singularity that corresponds to the string source at the location $\vec F(\vec x)$. 

In typical microstates, expectation values of $n_k$ follow the Bose/Fermi distributions; this implies that  $n_k\sim 1$ for generic $k$ (i.e.,~for wavelengths around the thermal value). Approaching such a typical microstate from states with larger $n_k$, one finds that the gravity solution does not tend towards the traditional black hole with a smooth horizon; rather, it tends to a quantum bound state with radius of order the horizon size. The study of this system led to the fuzzball conjecture: the size of bound states in string theory gives rise to quantum effects on the scale of the would-be horizon, so that no state in string theory produces a gravitational solution with horizon. More precisely, one never obtains a horizon with a neighborhood of spacetime around it where the dynamics of low energy modes is approximated by dynamics in the (say Unruh) vacuum; in particular the dynamics of the modes involved in Hawking radiation will differ {\it by order unity} from evolution in the  vacuum ~\cite{Lunin:2001jy,Mathur:2005zp,Bena:2007kg,Skenderis:2008qn,Mathur:2009hf,Mathur:2012zp}.\footnote{The black hole solution with horizon may still provide an excellent approximation for certain physical processes such as the infall of energetic matter onto the fuzzball~\cite{Mathur:2012jk,Mathur:2013gua}, or (for example) the emission of gravitational waves in black hole mergers~\cite{Abbott:2016blz}. In the emission of gravitational waves from a black hole merger, the key property of the horizon is to provide a purely ingoing boundary condition; such a condition is reproduced to a very good approximation by the highly absorptive fuzzball surface \cite{Guo:2017jmi}.}

The computation of \cite{Lunin:2001jy} considered vibrations of the string in the non-compact directions. 
Vibrations in the compact directions were studied in \cite{Lunin:2002iz,Kanitscheider:2007wq}, fermionic condensates were studied in~\cite{Taylor:2005db}, and vibrations of the heterotic string were studied in \cite{Kanitscheider:2007wq}. In each case no horizon or singularity was found, apart from the singularity corresponding to the string source, which is understood to be a physical source in the theory. 

By a sequence of S and T dualities, the fundamental string (NS1) and momentum (P) charges can be mapped to  D5 and D1 charges respectively~\cite{Martinec:1999sa,Lunin:2002iz,Mathur:2005zp,Kanitscheider:2007wq}. The supergravity solutions for supersymmetric \nsonep states are mapped under these dualities to supergravity solutions describing supersymmetric \donedfive states. These solutions are given in (\ref{D1D5Chiral}) below. The NS1 string source is mapped to a KK monopole supertube, which is a smooth gravitational solution with a topological twist. In the \donedfive duality frame the solutions are smooth, with no horizon or singularity of either kind: neither a black hole singularity nor a singularity arising from a string/brane source~\cite{Lunin:2002iz}. These smooth solutions were quantized and counted in~\cite{Rychkov:2005ji,Krishnan:2015vha}, and the microscopic entropy (\ref{one}) was reproduced. The D1-D5 solutions have been studied using precision holography~\cite{Kanitscheider:2006zf,Skenderis:2006ah,Taylor:2007hs,Giusto:2015dfa}. In the NS1-NS5 duality frame, some of these backgrounds have recently been  generalized to exactly-solvable worldsheet sigma models~\cite{Martinec:2017ztd}, enabling stringy aspects of black hole microstates to be explored~\cite{Martinec:2015pfa,Martinec:2018nco}.

The fuzzball paradigm would resolve the Information Paradox~\cite{Hawking:1976ra,Mathur:2009hf}: rather than radiate by pair production from the empty space around a horizon, a non-extremal fuzzball would unitarily radiate the information at its surface encoded in its structure. This unitary Hawking radiation has been studied in certain examples~\cite{Chowdhury:2007jx,Avery:2009xr,Chakrabarty:2015foa,Martinec:2018nco}.

In 2009 Sen made a proposal for how to deal with smooth classical solutions, for the purpose of microstate counting~\cite{Sen:2009bm}.  Sen argued that the smooth \donedfive solutions (\ref{D1D5Chiral}) should be regarded as a good classical description of the \donedfive system, while the dual \nsonep solutions (\ref{eq:ns1p-sol}) should not. Sen reaches this conclusion by working with a particular notion of a `classical solution'. Such a solution of string (field) theory should include corrections to all orders in $\alpha'$, but no corrections in the string coupling $g_s$. Sen argues that such a classical solution should not have string theory sources; it should be described by a smooth set of fields. With this definition, he observes that the \donedfive solutions (\ref{D1D5Chiral}) are good classical solutions while the \nsonep ones are not, since they contain a string source. He also argues that the traditional black hole with horizon should be considered a good `classical solution' for the \nsonep frame.\footnote{Note that in Sen's approach one counts a discrete set of quantum states, so the computation is not classical in this sense. The point is rather that one does not include corrections that are higher orders of the string coupling $g_s$, so the effective action will not include string loop corrections.} He concludes that in the \donedfive duality frame the entropy should be obtained by counting  solutions with no horizon, while in the \nsonep duality frame the entropy should be obtained from the Wald entropy computation for the geometry with horizon.

This approach may lead one to believe that in the \donedfive frame the microstates are fuzzballs, while in the \nsonep frame they are not;  i.e., that in the \nsonep frame one just has the traditional solution with horizon. We will argue that such an inference is incorrect, and that the microstates are fuzzballs in all duality frames. 
Note that the term fuzzball refers to general quantum bound states of string theory corresponding to black hole microstates; a given fuzzball state may or may not have a description as a classical solution, with or without sources. We expect typical fuzzball states to have Planck-scale degrees of freedom, meaning that typical states are not individually described by classical solutions. Having said this, families of supergravity solutions can give valuable insight into the structure of bound states as one takes the limit from atypical to typical states~\cite{Lunin:2001jy,Lunin:2002qf}.

The issue we address is important when it comes to more general black holes made from three or more charges. If we assume that  Sen's method of counting directly says something about the structure of microstates, then it would seem that one class of the microstates for such holes will be described through a traditional horizon, while another class, typically smaller in number, would be described by smooth solutions.

By contrast, the fuzzball paradigm proposes that no solution has a horizon. More precisely, one must examine the dynamics of the modes involved in radiation from the hole, and for this purpose both $\alpha'$ and $g_s$ effects must be considered whenever they are relevant. The traditional picture of the hole has these modes evolving in a local vacuum around the horizon, while in the fuzzball paradigm the state near the surface of the fuzzball has very little overlap with the vacuum state.

In this paper we will show that using Sen's argument to understand microstates faces two difficulties. These difficulties have been noted in passing elsewhere; in the present work we develop and expand upon these difficulties more explicitly, adding some simple computations to support the arguments. We then argue that Sen's proposal does not impact upon the fuzzball paradigm, and we instead interpret his approach as a string theory generalization of the Euclidean derivation of black hole entropy carried out by Gibbons and Hawking~\cite{Gibbons:1976ue}. 

%
A more detailed outline of these points is as follows:
%

(i) The proposal in \cite{Sen:2009bm} is based on a scaling 
argument. Sen considers the classical supergravity action 
which contains corrections to all orders in $\alpha'$ but 
not in $g_s$; solutions that extremize this action and that 
do not have string-theory sources are considered to be good 
`classical solutions'.  He notes that this action has a 
certain scaling symmetry in the fields and charges. 
From this scaling he shows that in the \donedfive duality
 frame, the area of the horizon of a putative classical 
black hole solution would not scale correctly with the 
charges; thus such a solution cannot exist and therefore 
the entropy cannot be described in terms of the area of 
such a horizon. However one can obtain the entropy in a 
different way: since the microstates (\ref{D1D5Chiral}) 
are smooth classical solutions, one can count them directly 
by quantizing the space of smooth solutions, and thereby 
obtain the entropy. In the \nsonep frame, on the other hand, 
the solutions (\ref{eq:ns1p-sol}) have a string source, 
and Sen would not consider these as valid classical 
solutions. However this time the putative horizon area 
scales correctly with the charges, and indeed the entropy 
can be obtained from such a black hole solution as 
mentioned above~\cite{Dabholkar:2004yr}.

One difficulty with this argument can be seen by 
considering a third duality frame: one in which the 
charges are \donep, as noted 
in~\cite{Black:2010uq}.\footnote{More generally, one 
can consider any duality frame in which the bound state 
has both R-R and NS-NS charges.} 
We write the corresponding microstate solutions 
in (\ref{D1Psol}) below; as expected these have a 
singularity corresponding to the D1-brane source.  
Thus these solutions should not be valid 
`classical solutions' according to the reasoning 
of~\cite{Sen:2009bm}, and so in Sen's framework they 
cannot be counted to obtain the entropy. 
However the putative horizon area does not scale 
correctly with the charges in this duality frame, 
so one cannot use a solution with horizon to obtain 
the entropy either. 
This indicates that the correct perspective is the 
fuzzball one, in which the two-charge solutions 
(\ref{eq:ns1p-sol}), (\ref{D1D5Chiral}), 
(\ref{D1Psol}) describe the microstates in the 
respective \nsonep, \donep, and \donedfive duality 
frames, and in no case does any microstate have a horizon. 
%

(ii) A second difficulty is as follows. In the \nsonep frame Sen appears to assume that the traditional supersymmetric small black hole solution with horizon describes a good string solution to all orders in $\alpha'$. A similar assumption is made for extremal black holes with more charges, where the horizon has a macroscopic size (i.e.~a radius much larger than the string length $\ell_s=\sqrt{\alpha'}$). In both cases, the extremal black holes have a throat which has infinite proper length when measured along a spacelike slice. It may  therefore seem that the horizon is infinitely far away, and the geometry outside the horizon is a complete solution to the tree-level string action.

However it was noted in \cite{Mathur:2014nja} that this is not the case. As is well known, while the proper length of the throat measured along a spacelike slice is infinite, freely-falling massive particles fall across the horizon in finite proper time. As we will describe in detail below, there exist finite-action paths that connect points in the throat to the singularity. Thus in the string worldsheet dynamics in the throat, one is forced to consider tubes that emerge from the worldsheet and touch the singularity. One must thus have a consistent boundary condition at the singularity.

In string theory one allows certain singularities that correspond to 1/2-BPS brane sources. By contrast, in the traditional geometry of a two-charge BPS black hole, the singularity carries two charges, and one does not have a boundary condition for such a source. The fuzzball construction demonstrates that such two-charge singularities resolve themselves into a fuzzball structure where locally the singularity is due to a {\it single} brane charge (e.g.~a D1-brane at an angle, as a local segment of a D1-P fuzzball). We indeed have valid boundary conditions for such a resolved source. Thus we conclude that the spherically symmetric black hole solution considered by Sen is not a valid solution to the string dynamics, and correcting it to a valid solution by resolving the singularity in string theory brings us back to a fuzzball solution with no horizon, regardless of which duality frame we are working in.
%
%

(iii) In other works, Sen and others have used the traditional solution with horizon to good effect in computing the entropy for extremal black holes (see e.g.~the review~\cite{Sen:2007qy}). If the classical solution is invalid for the above reason, then how should one interpret the success of these entropy computations? As noted in \cite{Mathur:2014nja}
one should Wick rotate the Lorentzian black hole by $t\r -i\tau$ to obtain a Euclidean solution. This Euclidean solution is a saddle point in the path integral over all gravity solutions, and expanding about this saddle point should give the entropy the same way that Gibbons and Hawking obtained the leading order entropy in \cite{Gibbons:1976ue} (c.f.~the related remarks in~\cite{Skenderis:2008qn}). Sen's computations should be unaffected by this rotation, and the microscopic counts should be reproduced again. The distinction we make is about the physics of the original Lorentzian microstates: we argue that these should be  fuzzballs in each duality frame, with no horizon. The Euclidean saddle point approximates a path integral where all these fuzzball states run in a $\tau$ loop, but we should not Wick rotate this saddle point  back to Lorentzian signature and argue that it gives a  valid string solution. 
%

As this paper was in an advanced stage of preparation, we received 
the preprint \cite{Cano:2018hut} which notes that for the (4+1)-dimensional
two-charge extremal black hole created by the heterotic string carrying momentum, 
the numerical coefficient in the Wald entropy does not agree with the 
numerical factor in the microscopic entropy. 
This supports the view that one should compute the entropy 
of this system by counting fuzzball states, regardless of the duality frame.

The remainder of this paper is organized as follows. In Section \ref{sec:sugra-solns} we review the two-charge supergravity solutions in different duality frames. In Section \ref{sec:scaling-arg} we review Sen's scaling argument and discuss its problems. In Section \ref{sec:incomplete-string-backgd} we argue that the traditional black hole solution with horizon is not a valid string background because of its unphysical singularity. In Section \ref{sec:trad-sol-interpretation} we argue that the traditional black hole solution with horizon is best understood through its continuation to a smooth Euclidean solution. Section \ref{sec:disc} contains further discussion.


\section{The two-charge supergravity solutions} \label{sec:sugra-solns}

Consider Type IIB string theory compactified to $M_{4,1}\times S^1\times T^4$, as in Eq.\;(\ref{eq:compactification}). Let the $S^1$ be parametrized by a coordinate $y$ with
\be
0 \, \le \, y \, < \, L \,.
\ee
Let the time coordinate be $t$ and the non-compact spatial directions of $M_{4,1}$ be $x_1, \dots, x_4$. 
Let the $T^4$ have a rectangular shape, with coordinate volume $V$, parametrized by $z_1, \dots, z_4$.

Consider states with $n_w$ units of NS1 string winding around $S^1$, and $n_p$ units of momentum along $S^1$. The bound state with these charges is a multi-wound fundamental string that is wound $n_w$ times around the $S^1$ and carries $n_p$ units of momentum charge in the form of travelling waves. 

It is convenient to consider the $n_w$-fold covering space of the $S^1$. In this covering space the multi-wound string is a singly-wound string of total length $L_T \equiv n_w L$. In the classical limit of large $n_w, n_p$, coherent states are described by a classical vibration profile for the travelling wave $\vec F(\hat{v})$, where $\hat{v}=t-\hat{y}$ and where $\hat{y}$ is the covering-space coordinate, $0 \le \hat{y} < L_T$. For simplicity we restrict to $\vec F$ having non-zero components only along the four non-compact directions $x_1, \dots ,x_4$. 

In spacetime (as distinct from the covering space) this configuration can be described as $n_w$ strands of the NS1 with appropriate joining conditions, and with each strand carrying a transverse vibration. It is possible to write the metric for a string carrying a travelling wave, and further, to superpose such strings when the momentum on each is carried in the same direction $y$. In the limit of large $n_w$, one can smear over the strands and describe them as a continuous distribution. 

The resulting \nsonep family of supergravity solutions, in the string frame, is given by
\bea
ds^2 &=&{1\over H}\Big( \!\!\: -dudv+Kdv^2+2A_i \:\! dx_i dv\Big)+\sum_{i=1}^4 dx_idx_i+\sum_{a=1}^4 dz_adz_a \,, \nn
e^{2\Phi}&=&H^{-1} \,, ~~\qquad B_{uv}~=~ -{1\over 2}[H^{-1}-1] \,,  ~~\qquad B_{vi}=H^{-1}A_i \,, 
\label{eq:ns1p-sol} 
\eea
where
\be \label{eq:F1P-HAK}
H \,=\, 1+{Q_1\over L_T}\int\limits_0^{L_T}  {d\hat v\over |\vec x-\vec F(\hat v)|^2} \;, \quad
A_i \,=\, -{Q_1\over L_T}\int\limits_0^{L_T}  {d\hat v \, \dot F_i(\hat v)\over |\vec x-\vec F(\hat v)|^2} \;, \quad
K \,=\, {Q_1\over L_T}\int\limits_0^{L_T} {d\hat v \, |\dot
F(\hat v)|^2\over |\vec x-\vec F(\hat v)|^2} \;.
\ee
Here $F_i(\hat{v}+L_T)=F_i(\hat{v})$ and  $\dot F$ denotes the derivative of $F$ with respect to
$\hat{v}$.\footnote{The solutions described above are constructed at a special point in 
the moduli space of compactifications. One can make solutions at more 
general moduli by taking, for instance, the compact directions 
$S^1\times T^4$ to form a $T^5$ which is not rectangular. 
One still finds fuzzball solutions (rather than a black hole with horizon). 
The only new feature is that (for coprime $n_1, n_p$) the solution cannot 
break into two parts with no cost in energy; 
i.e., the solution is bound rather than threshold bound.}

One can perform the following dualities on these solutions to map them to the \donedfive solutions in Type IIB:
\be
\begin{array}{c} 
{\rm NS1}_y \\
{\rm P}_y \end{array}
~~ \stackrel{{\rm S}}{\r}~~
\begin{array}{c} 
{\rm D1}_y \\
{\rm P}_y \end{array}
~~ \stackrel{{\rm T}_{5678}\phantom{\big]}}{\r} ~~
\begin{array}{c} 
{\rm D5}_{y5678} \\
{\rm P}_y \end{array}
~~ \stackrel{{\rm S}}{\r} ~~
\begin{array}{c} 
{\rm NS5}_{y5678} \\
{\rm P}_y \end{array}
~~ \stackrel{{\rm T_{y6}}}{\r} ~~
\begin{array}{c} 
{\rm NS5}_{y5678} \\
{\rm F1}_y \end{array}
~~ \stackrel{{\rm S}}{\r} ~~
\begin{array}{c} 
{\rm D5}_{y5678} \\
{\rm D1}_y \end{array}
\ee
Performing these dualities on the \nsonep solutions (\ref{eq:ns1p-sol}) using the Buscher rules, one obtains the \donedfive string-frame solutions:
\bea\label{D1D5Chiral}
ds^2 &=&\sqrt{\frac{H}{1+K}}\left[-(dt-A_idx^i)^2+(dy+B_idx^i)^2\right]
+
\sqrt{\frac{1+K}{H}} \;\! d{\vec x}\cdot d{\vec x}\nonumber\\
&&{} + \sqrt{H(1+K)}\, d{\vec z}\cdot d{\vec z} \,, \\
e^{2\Phi}&=&H(1+K) \,,\qquad
C^{(2)}_{ti}=\frac{B_i}{1+K} \,,\qquad
C^{(2)}_{ty}=-\frac{K}{1+K} \,,\nonumber\\
C^{(2)}_{iy}&=&-\frac{A_i}{1+K} \,,\qquad
C^{(2)}_{ij}=C_{ij}+\frac{A_iB_j-A_jB_i}{1+K} \;,
\eea
where the forms $B_i$ and $C_{ij}$ are defined by the duality relations:
\be\label{DualFields}
dB=-\ast dA, \qquad dC=-\ast d(H^{-1})
\ee
and the harmonic functions $H, K, A_i$ take the same form as in (\ref{eq:F1P-HAK}), up to scalings of the coupling constants etc., that arise from the duality maps. (These scalings are given in \cite{Mathur:2005zp}.)

In the set of dualities we performed, after the first S-duality one obtains the charges \donep; the string-frame supergravity fields for this configuration are
\bea\label{D1Psol}
ds^2 &=&   H^{-\frac{1}{2}} dv \;\Big(\!\! - du +  K  dv +
2 A_{i}  dx^i \Big) +  H^{\frac{1}{2}} dx^i dx^i \,,  \\ 
e^{2 \Phi} &=&  g_s^2 H \,, \qquad
C^{(2)}_{uv} ~=~ - \tha (H^{-1} -1) \,, \qquad
C^{(2)}_{vi} ~=~ - H^{-1} A_i \;. \nonumber
\eea
Again the harmonic functions are given by appropriately transformed versions of the functions written in~\eq{eq:F1P-HAK} \cite{Mathur:2005zp}.

\section{The scaling argument}
\label{sec:scaling-arg} 

In this section we first review some of the reasoning behind Sen's suggestion to treat smooth source-free solutions differently from solutions with sources. We then review Sen's scaling argument that in different duality frames the entropy of two-charge microstates should be accounted for in different ways: by counting solutions without sources or by the entropy associated to a horizon, and thus that more generally one should add the contributions from smooth source-free horizonless solutions and solutions with horizons. We then observe that such an argument does not allow either method of counting to be used for the \donep duality frame, or any other frame with mixed R-R and NS-NS charges.

\subsection{Supergravity solutions without sources}

As we have noted, the fuzzball paradigm would resolve the information paradox: a non-extremal fuzzball would radiate the information at its surface encoded in its structure, rather than radiate by pair production from a horizon. However it is interesting to ask more detailed questions about the nature of the fuzzball microstates, and in particular about the role of smooth source-free supergravity solutions.

A genuine complication of dealing with smooth source-free solutions arises in the context of precision counting of black hole microstates~\cite{Banerjee:2009uk,Jatkar:2009yd}. Given a brane bound state corresponding to an extremal black hole, the vast majority of microstates reside deep inside the throat region of the spacetime, where the traditional geometry has a horizon. However a small number of microstates can arise at the region at the top of the throat, where it changes over to flat spacetime; this region is often referred to as the `neck' region. These states are described by smooth deformations of the supergravity solution at the neck, with no string sources present, and we will denote them by `neck modes'.\footnote{These states were described by Sen as `hair' in a somewhat unfortunate terminology; in earlier literature, `hair' refers to degrees of freedom at the horizon of a black hole (see e.g.~the review~\cite{Chrusciel:2012jk}).}
The entropy of these modes must be added to the entropy of the black hole in order to correctly reproduce the microscopic entropy~\cite{Banerjee:2009uk,Jatkar:2009yd}.  
One can ask more generally: how many microstates of a given black hole are accounted for by quantizing the space of smooth supergravity solutions?

In the fuzzball paradigm the horizon is replaced by a quantum bound state of string theory. When this bound state can be described geometrically, the solution describing the bound state caps off smoothly deep inside the throat region. The details of the solution near the cap encode the information of the microstate. However one can also consider neck modes that modify a given fuzzball state at the top of the throat. Such neck modes can be constructed explicitly, are globally smooth, and have a consistent holographic interpretation~\cite{Mathur:2011gz,Mathur:2012tj,Lunin:2012gp,Giusto:2013bda,Mishra:2018bcb}.\footnote{We note however that these neck modes are given by a different solution to the field equations than the one selected in~\cite{Jatkar:2009yd}. The supergravity equation is a second order differential equation with two solutions, and  the selected branch is different in the two cases.}

Because of the existence of neck modes, we see that the entropy of at least {\it some} smooth classical field profiles must be added to the entropy that arises from the degrees of freedom deep inside the throat, where the horizon would be. 
It is natural to explore what general lesson, or rule, one should derive from this. 
For the two-charge black hole, Sen noted that in the \donedfive frame the would-be horizon is replaced by a family of smooth source-free capped solutions (and limits thereof). He thus suggested that these smooth solutions should be quantized and counted, just as must be done for the neck modes. By contrast, in the \nsonep frame Sen noted that, at least for heterotic compactifications, one can obtain (after stringy corrections) a solution with a horizon. Thus one would double-count if one added the entropy of the smooth solutions with the string source to the entropy associated with the horizon. Sen proposed not to count the microstates with the string source, and instead to obtain the count of microstates in this duality frame from the area of the horizon of the $\alpha'$-corrected black hole solution. 

While there may be different ways to count microstates, we emphasize that our primary interest is in a different question: what is the spacetime structure of the black hole microstates? One might use Sen's counting approach to argue that some states are explicit string theory solutions while others can only be represented as an ensemble through a horizon. We will argue that this would be incorrect, and that the two-charge microstates are fuzzballs in all duality frames, regardless of whether there may be sources in some frames and not in others. We will furthermore argue that the solution with a horizon is not a valid string solution in any duality frame.

We next proceed to review Sen's scaling argument which seeks to distinguish the \donedfive frame from the \nsonep frame.

\subsection{The scaling argument}

The bosonic part of the action of Type IIB supergravity is
\bea
S_{\textsc{IIB}}&=&{1\over 2\kappa^2}\int d^{10} x \sqrt{-g} 
\left [ e^{-2\phi} \left( R +4 \:\! \p\phi \:\! \p\phi-\h H^2 \right) - \left( \h F_1^2+\h F_3^2+{1\over 4} F_5^2 \right) \right ]\nn
&~~~&-{1\over 4\kappa^2}\int C_4\wedge H_3\wedge F_3 \,,
\eea
with a self-duality constraint on $F_5$. Here $H=dB_2$, $F_3=dC_2-C_0 H_3$, etc. 

Sen observes that this action has the following scaling symmetry~\cite{Sen:2009bm}:
\bea
\phi&\r& \phi+\ln \lambda^{-1} ~~~\implies~~~e^{-2\phi}\r \lambda^2\, e^{-2\phi} \,,\nn
g_{\mu\nu}&\r& g_{\mu\nu}\,,\nn
B_2&\r& B_2 \,,\\
C_n&\r&\lambda C_n ~~~\implies~~~F_{n+1}\r \lambda F_{n+1} \,.\nonumber
\eea
Under these scalings we obtain
\bea
S_{\textsc{IIB}}&\r& \lambda^2 S_{\textsc{IIB}} \,.
\label{twonew}
\eea
Now let us consider the charges. Magnetic charges are given by integrating the corresponding field strength over a Gaussian surface.  We thus obtain the scalings
\fontdimen17\textfont2=5pt
\bea
q^{\mag}_{\NSNS}~\r~ q^{\mag}_{\NSNS} \,, \qquad 
q^{\mag}_{\rm{RR}}~\r~ \lambda \;\! q^{\mag}_{\rm{RR}} \,.
\eea
\fontdimen17\textfont2=2.5pt
On the other hand for an electric charge such as that of the fundamental string, we have the equation
\be
\left ( e^{-2\phi} H_{\mu\nu\lambda}\right ) ^{;\lambda}\;=\;j^{\el}_{\mu\nu} \,.
\ee
If we keep the field $B_2$ unchanged, while scaling $e^{-2\phi}\r \lambda^2 e^{-2\phi}$, then we must scale $j^{\el}$ by $\lambda^2$. For an RR field, we have a field equation of the form
\be
\left (  F_{\mu\nu\lambda}\right ) ^{;\lambda}=j^{\el}_{\mu\nu} \,.
\ee
Thus if we scale $C_2\r\lambda \;\!  C_2$, then we must scale $j^{\el}_{\mu\nu}\r \lambda \;\!  j^{\el}_{\mu\nu}$.
With this reasoning,  we find
\fontdimen17\textfont2=5pt
\bea
q^{\el}_{\NSNS}~\r~\lambda^2 q^{\el}_{\NSNS}  \,, \qquad 
q^{\el}_{\RR}~\r~ \lambda \;\!  q^{\el}_{\RR} \,.
\eea
\fontdimen17\textfont2=2.5pt
The entropy of the black hole can be derived from the action computed for a solution of the field equations, and is found to be proportional to the on-shell value of the action.\footnote{For non-extremal black holes this can be seen from the Gibbons-Hawking computation using the Euclidean black hole. For extremal black holes, one must start with the corresponding non-extremal black hole and then take the zero-temperature limit.} For extremal black holes the entropy depends only on the charges. Thus from (\ref{twonew}) and the scalings of the charges we find the scaling relation
\fontdimen17\textfont2=5pt
\bea
S_{\BH} \! \left( \lambda^2 q^{\el}_{\NSNS}, \;\! \lambda \:\! q^{\el}_{\RR}, \;\!q^{\mag}_{\NSNS}, \;\!\lambda \:\! q^{\mag}_{\RR}\right) 
&=& \lambda^2 S_{\BH} \! \left( q^{\el}_{\NSNS}, \;\! q^{\el}_{\RR}, \;\!  q^{\mag}_{\NSNS}, \;\! q^{\mag}_{\RR}\right).
\label{three}
\eea
\fontdimen17\textfont2=2.5pt
While we worked with Type IIB supergravity above, a similar argument applies in Type IIA, and also in the heterotic theory where we of course have NSNS fields but no RR fields.

We now examine the argument of \cite{Sen:2009bm}. Consider the supersymmetric \nsonep black hole solution in IIA/IIB, or the supersymmetric heterotic string solution carrying string winding and momentum charges. These charges are all electric. The leading-order solution  produced by these charges does not have a horizon; it has a singularity at $r=0$. However if the curvature becomes string scale at some location $r_h$, there will be corrections to the metric from terms in the effective action that are higher order in $\alpha'$. Such $\alpha'$ corrections may arise through curvature invariants such as
\be
R_{\mu\nu\lambda\kappa}R^{\mu\nu\lambda\kappa}(r_h)\sim {1\over \alpha'^2} ~,
\ee
or alternately may arise through null components of the ``T tensors'' formed from curvature and field strength tensors, even if they do not arise in any coordinate invariants, as has been recently pointed out~\cite{Cano:2018qev,Cano:2018aod}.
Let us assume that these $\alpha'$ corrections change the geometry to that of an extremal hole with a horizon area equal to the surface area at the location $r_h$. For such a corrected geometry we can compute the on-shell action $S$ and thereby an entropy $S_{\BH}$. 

Now suppose we scale the charges as in (\ref{three}). The metric $g_{\mu\nu}$ is not scaled, so the geometry (with all its $\alpha'$ corrections) remains unchanged. The relation (\ref{three})  then gives
\bea
S_{\BH}(\lambda^2 n_w , \:\! \lambda^2 n_p)&=&\lambda^2 S_{\BH}(n_w,  n_p) \,.
\eea
This scaling is reproduced by the corresponding scaling of the microscopic entropy,
\be
S_{\micro}(\lambda^2 n_w ,\:\! \lambda^2 n_p)=2\pi\left (\lambda^2 n_w , \:\! \lambda^2 n_p \right ) ^\h  =2\pi\lambda^2 (n_w, n_p)^\h=\lambda^2 S_{\micro}(n_w, n_p) \,.
\ee
From this discussion we conclude that an $\alpha'$-corrected black hole geometry would be compatible with the scaling of the microscopic entropy. (Note however, that such an  argument does not {\it prove} that a spherically symmetric extremal black hole is indeed the correct geometry to describe the \nsonep extremal hole, and we will argue later that it is not.)

Sen then examines the \donedfive system, which is obtained from \nsonep by dualities. Both the D1 and D5 charges are RR charges. Thus if the entropy for this system could be obtained from a  classical solution, we would find the scaling
\bea
S_{\BH}(\lambda n_1 , \:\! \lambda n_5)&=&\lambda^2 S_{\BH}(n_1,  n_5) \,.
\eea
However the microscopic entropy $S_{\micro}=2\pi\sqrt{n_1n_5}$ does not satisfy this scaling:
\be
S_{\micro}(\lambda n_1, \:\! \lambda n_5)
 ~=~2\pi\lambda \sqrt{n_1n_5}~=~\lambda \:\! S_{\micro}(n_1,  n_5)\,.
\ee
From this Sen concludes that in the \donedfive duality frame, a black hole solution cannot be used to derive the entropy of the system, and the entropy must therefore be accounted for by quantizing the space of smooth fuzzball solutions (\ref{D1D5Chiral}). 

For the \nsonep system, however, Sen argues that one should not quantize the space of solutions (\ref{eq:ns1p-sol}) and then count their number. His reason is that these solutions have a singularity arising from the NS1 string source. This makes the solutions singular, and such solutions should not be considered valid `classical solutions' for the purpose of counting. As we have seen, in this case one can in principle obtain the entropy from a black hole solution instead, since the scaling law works in the \nsonep frame. 

\subsection{Problem with the scaling argument}\label{secpt}

An immediate problem with the argument reviewed above is as follows. Consider the two-charge supersymmetric solutions in the 
\donep duality frame, as given in (\ref{D1Psol}). These solutions have a D1-brane source, so in Sen's framework they would not be considered valid classical solutions. Thus in this framework the entropy cannot be obtained by  quantizing the space of these solutions and then counting them. In that case we should look for the entropy from a black hole solution. However we find that in this frame the scaling indicates that there cannot be a classical black hole solution. The D1 charge is an RR charge while P is an NSNS charge. The scaling relation (\ref{three}) gives
\bea
S_{\BH}(\lambda n_1 , \:\! \lambda^2 n_p)&=&\lambda^2 S_{\BH}(n_1, n_5)
\eea
while the microscopic entropy $S_{\micro}=2\pi\sqrt{n_1n_p}$ scales as
\be
S_{\micro}(\lambda n_1, \:\! \lambda^2 n_p)
 ~=~2\pi\lambda^{3\over 2} \sqrt{n_1n_5}
~=~\lambda^{3\over 2} S_{\micro}(n_1, n_p)\,.
\ee
Thus we find that Sen's framework does not account for 
the entropy of the 
\donep system; neither smooth source-free solutions 
nor a black hole geometry 
appear to account for the entropy. This is one of 
the difficulties that we find with Sen's scaling 
argument.

It is possible that this sharp contradiction could 
be avoided if there 
were some other smooth solutions without sources 
describing the D1-P bound states; 
however the computation of~\cite{Black:2010uq} appears to make this possibility 
rather unlikely.

One might also try to avoid the above contradiction by
proposing that the D1-P solutions be considered 
to be valid classical solutions, 
since the D1 source corresponds to a nonperturbative 
object in string theory, 
and thus may be considered to be closer to a monopole rather than a 
fundamental string.\footnote{We thank A.~Sen for suggesting this possibility.}
One might still propose that the NS1-P system is 
described by the black hole solution. 
There is a difficulty, however, with such a proposal. 
Let us work in IIB string theory and let us first 
consider both the NS1-P and D1-P 
bound states at string coupling $g_s\ll 1$.
In this proposal, the NS1-P bound state should be 
described by a black hole while the D1-P bound state 
is described by fuzzball solutions. 
Now let us continuously change $g_s$ to a value $g_s\gg1$. Then the NS1 and D1 interchange 
roles; in fact we can interchange them by an S-duality to return to 
small $g_s$ configurations. Thus as we increase $g_s$ in this scenario, at some point the 
fuzzballs of the D1-P system would need to transition to being described by a black hole. 
This appears strange, since one expects a continuous change of 
the wavefunction as we change $g_s$.
The fuzzballs have a finite-length throat, and a change of coupling 
can change the length of this throat. However the black hole has an 
{\it infinite} throat, and it is very unclear how a continuous change 
of wavefunction could cause a transition from finite throats to an infinite throat, and vice versa.

\subsection{Proposed resolution}
\label{sec:local-dipole-charges}

In the fuzzball paradigm, the microstates in all 
duality frames (\nsonep, \donedfive, \donep etc) 
describe the actual microstates of the black hole, 
and should be counted (after appropriate quantization) 
to obtain the entropy of the system. 
The black hole solution with horizon is not a 
correct Lorentzian solution for any duality frame. 
Having said this, we will discuss in 
Section~\ref{sec:trad-sol-interpretation} how  
Euclidean (rather than Lorentzian) black hole 
solutions can be used as very useful technical 
tools for computing black hole entropy. 

For the purpose of investigating the spacetime 
structure of the bound states, it is not 
particularly important whether the solution 
is sourceless (as in the \donedfive frame) or 
has a source (as in the \nsonep and \donep frames). 
Indeed, in string theory there is a simple relation 
between sources and smooth solutions. 
Any single charge like an NS1 string can be 
dualized to any other charge (e.g.~P, NS5, D$p$, etc). 
In particular it can be dualized to a KK monopole  
(by dualizing to D6 and then using a 9-11 flip). 
The KK monopole is a solution of gravity with no 
source, and is indeed a smooth solution when 
$x_{11}$ has a large radius. 
Dualities of course modify the couplings and moduli of the 
theory, but our emphasis here is on whether an 
object has a `source' or not. In this 
language the fundamental string has a source, while the 
KK monopole, to which it can be mapped, does not.  
However from the existence of such duality maps 
we see that there is no fundamental difference 
between smooth solutions and solutions with 
allowed sources. Here the term `allowed' is important: 
sources give a singularity in the supergravity fields, 
but the singularities that are `allowed' are only those 
singularities that have a physical interpretation 
in string theory; other singularities would not 
be valid solutions of the theory.  

The previous comment on the smoothness of KK monopole charge is directly relevant to the issue that Sen discusses, so is worth discussing in more detail. Consider an \nsonep bound state. The winding and momentum charges are along the compact $S^1$ direction $y$. However the momentum is carried by transverse vibrations. Consider  a point on the string, and suppose that at this point the vector $\dot F_i(v)$ is locally purely along the non-compact direction $x^1$. Then locally we have a segment of fundamental string that is a straight line oriented along a direction 
\be
\hat n = \cos\theta \, {\hat n}_y +\sin\theta \, {\hat n}_1
\ee
for some angle $\theta$, where ${\hat n}_y$, ${\hat n}_1$ denote unit vectors along $y$ and $x^1$ respectively. The string can only carry momentum transverse to itself, so the momentum direction is along
\be
\hat n'= \sin\theta \, {\hat n}_y -\cos\theta \, {\hat n}_1 \,.
\ee
The components along ${\hat n}_y$ sum up to give the conserved winding charge NS1$_y$ and momentum charge P$_y$ of the configuration, while the components along ${\hat n}_1$ give rise to dipole charges NS1$_1$ and P$_1$, since at some other point along the string the motion has to be in the opposite direction along $\hat x^1$, to make the string close after $n_w$ turns, such that the net dipole NS1$_1$ charge is zero. Similarly the dipole P$_1$ charge adds to zero since the string has no overall momentum P$_1$.

On dualizing to the \donep frame, the dipole charges become D1$_1$, P$_1$. On further dualizing to \donedfive, the dipole charges become KK monopole and momentum P \cite{Lunin:2002iz}. This is why the \donedfive solutions are smooth. But this smoothness is simply a particular feature of this duality frame, arising from the fact that the KK monopole is a smooth fundamental charge; this dipole charge appears as a singular source if dualized to other frames.

We therefore argue that as far as the fuzzball nature of microstates is concerned,  there is nothing special about the \donedfive frame where the solutions appear smooth; in all duality frames the microstates should be given by the appropriate dualizations of the solutions (\ref{eq:ns1p-sol}), (\ref{D1D5Chiral}), (\ref{D1Psol}). Neck modes, when present, should be added to the entropy coming from the microstates that replace the horizon. The throat of the hole is parametrically long compared to its radius, so it is not difficult to distinguish modes localized near the neck from modes localized at the cap.

\subsection{The limit from non-generic to generic states}
\label{sec:limit-to-generic}

One of the difficulties with understanding the structure of black holes has been that the generic microstate is expected to have a complicated structure at the Planck scale, and this structure is difficult to describe explicitly. A key aspect of the fuzzball program is to manage this difficulty by taking a limit through non-generic microstates. One first uses the understanding of black hole entropy to 
list a basis of microstates in a sequence, starting from the simplest ones. Thus for \nsonep one knows that all states are given by vibrations of the NS1, and the simplest ones are those where all the momentum is carried by just one harmonic. In the next simplest ones we split the momentum among two harmonics, and so on. The generic state has excitation of order unity for each typical harmonic. For the simple microstates we can make explicit solutions using the techniques of string theory, since the large occupation number per mode makes the classical approximation a good one. This allows us to see the nature of the microstate, and in particular we see that  there is no horizon.

Similarly, for three-charge \donedfivep bound states, we write all states in terms of momentum excitations of an orbifold CFT having different twist sectors, and then construct simple microstates starting with the lowest twist sectors carrying all the momentum in only a few harmonics. Again no horizon has been found for any microstate that has been constructed. For recent progress on constructing and studying such solutions, see~\cite{Bena:2015bea,Bena:2016agb,Bena:2016ypk,Bena:2017geu,Bena:2017upb,Bena:2017xbt,Bena:2018bbd}.

One then examines the low-energy modes involved in Hawking radiation, and one observes that the dynamics of these modes in the simple microstates is not the dynamics in the traditional black hole geometry with horizon; it differs by order unity. As we take the limit to more complicated microstates, we see that the dynamics stays different  by order unity, instead of approaching the dynamics expected from a vacuum horizon. From this we get the fuzzball proposal for resolution of the information paradox: the state around the horizon is not close to the vacuum state for the purposes of the dynamics of the modes involved in Hawking radiation.

We can now see another difficulty that arises if one were to interpret Sen's counting prescription as implying that the fuzzball solutions (\ref{eq:ns1p-sol}) are not the microstates in the \nsonep frame. Consider the \donedfive solutions constructed in \cite{Balasubramanian:2000rt,Maldacena:2000dr} which, in the AdS decoupling limit, are $\mathbb{Z}_k$ orbifolds of AdS$_3\times S^3$. In these geometries one can compute the energy gap for excitations to the nearest non-extremal state. This computation was carried out in \cite{Lunin:2001jy}, and the result
\be
\Delta E={4\pi\over kL}
\label{five}
\ee
agrees with the expectation from the \donedfive CFT.  But now we can map these geometries to the \nsonep duality frame, as done in \cite{Lunin:2001fv}. We do so by transforming all moduli appropriately, and by dualizing both the geometry and the perturbation creating the non-extremal energy at the same time. This gives an equivalent dual description of the physics, so the resulting \nsonep microstate geometries (\ref{eq:ns1p-sol}) will give the same energy gaps (\ref{five}).  By contrast, if we assume that the \nsonep solution was described instead by the black hole, then we would find a {\it zero} energy gap at leading order.  Once we take into account backreaction we would find that for energies above a certain scale $E_{\mathrm{min}}$ the effect of backreaction cannot be ignored.\footnote{One of the authors (SDM) thanks A.~Sen for a discussion on this point.} But  this $E_{\mathrm{min}}$ does not depend on the choice of microstate (since it is based on the naive solution), and so does not agree with (\ref{five}). Thus we have the problem that if we replace all the states of the hole with one geometry -- the traditional hole -- then we cannot explain the different energy gaps that one finds for different states.

Thus at least for the simple microstate geometries with which we start the fuzzball analysis, 
we see that the solutions (\ref{eq:ns1p-sol}) describe the microstates in the \nsonep frame just as the solutions (\ref{D1D5Chiral}) describe the microstates in the \donedfive frame. 

Similarly, one may consider the scattering of low energy supergravity quanta off a  \donedfive microstate; this is explicitly computable for simple microstates. One will find that the details of the scattering depend on the choice of microstate. By duality, one would expect the same scattering off the \nsonep dual of this microstate. The corresponding \nsonep fuzzball (\ref{eq:ns1p-sol}) would indeed reproduce the scattering as desired, but if we try to use Sen's black hole solution for the \nsonep duality frame, then the scattering will be independent of the choice of microstate; further, some part of the wavefunction will be lost through the horizon. Thus we again see that the microstates in the \nsonep frame should be given by the fuzzball solutions (\ref{eq:ns1p-sol}).

It has been noted that in the study of two-charge microstates there are regimes in which the supergravity approximation no longer holds due to some degree(s) of freedom beyond the supergravity sector becoming light~\cite{Martinec:1999sa,Chen:2014loa}.\footnote{See also the remarks regarding the supergravity approximation in~\cite{Skenderis:2008qn}.} In that case the correct description is of course to incorporate correctly the relevant light degrees of freedom, which can be done either in the same duality frame or by changing duality frame. We re-emphasize that in all examples that have been studied, when taking the limit towards such regimes from regimes that are under full control, one always finds a fuzzball structure rather than empty spacetime around a horizon.

\section{Incompleteness of the throat geometry}
\label{sec:incomplete-string-backgd}

If one considers the spherically symmetric geometry obtained for the \nsonep system after $\alpha'$ corrections, then this geometry may be said to have a horizon, where the Wald entropy of this horizon reproduces the microscopic entropy. Similarly, if we consider extremal black holes with three or four charges, we can write spherically symmetric geometries which have a macroscopic horizon, from which an entropy can be computed. Since our arguments below will be very general, for concreteness and ease of notation we shall focus on the four-charge black hole in 3+1 noncompact dimensions; furthermore, for convenience we set the four charges to be equal in value, so that the (3+1)-dimensional solution is simply the Reissner-Nordstr\"om solution. Due to the generality of the arguments it is natural to expect that other extremal black holes, including the \nsonep hole, will display the same essential behavior.

The extremal Reissner-Nordstrom geometry has the 
following structure. 
Near asymptotic infinity we have flat spacetime, 
then we have a `neck' region, 
then a near-horizon throat 
that has infinite proper length as measured 
along a radial spacelike geodesic. 
There is a horizon at the end of this throat, 
and then a region inside the horizon in which 
there is a curvature singularity. 

Since the throat has infinite proper length as measured along a radial spacelike geodesic, it may appear that the exterior geometry from infinity up to the horizon is a complete spacetime, which after $\alpha'$ corrections may become a valid string background. We will argue that this is not the case: if one attempts to verify the vanishing of string worldsheet $\beta$-functions in this background, then the worldsheet fluctuations will explore the interior of the horizon. Furthermore, to obtain a complete string background one would need to specify boundary conditions at the singularity, and the black hole singularity does not give a valid set of boundary conditions.

\subsection{The extremal Reissner-Nordstrom geometry}

The metric of the extremal Reissner-Nordstrom solution in (3+1) dimensions is
\be
ds^2\;=\;-f(r) \:\! dt^2+{dr^2\over f(r)}+r^2(d\theta^2+\sin^2\theta d\phi^2) \,, \qquad\quad f(r) \;=\; \left(1-{M\over r}\right)^{\! 2} \,.
\label{qone}
\ee
Let us start by reviewing some properties of the 
extremal black hole solution. Consider a spacelike 
radial geodesic, at constant $t$, 
that starts at a point $r=r_0$ outside the horizon 
and ends at the horizon at $r=M$. 
The proper distance along this geodesic diverges:
\be
s\;=\;\int\limits_{r=M}^{r_0} {dr\over \left(1-{M\over r}\right)} ~\to~\infty \;.
\ee
While it may naively appear that the horizon is 
infinitely far away from $r=r_0$ and will thus 
never be reached by particles at $r=r_0$, 
it is a well-known fact that it takes a finite 
proper time for particles to fall radially 
from $r_0>M$ to the horizon at $r=M$. 
Consider a particle with energy per unit mass 
$\varepsilon$ at infinity, falling radially inwards 
in the geometry (\ref{qone}). Denoting the 4-velocity as $U^a={d\xi^a\over d\tau}$, we have that
\be
U_t\;=\;g_{tt} U^t\;=\;-\left(1-{M\over r}\right)^{\! 2} U^t \;\equiv\; -\varepsilon 
\ee
 is conserved along the infall. Using $U^aU_a=-1$, the proper time for infall from a radius $r_0$ to the horizon at $r=M$ is
 \be
 \tau(r_0)\;=\;\int\limits_{r=M}^{r_0} {dr\, r\over \sqrt{(\varepsilon^2-1)r^2+2Mr-M^2}} \;.
\ee
In particular, for a particle with $\varepsilon=1$ (e.g.~a particle falling from rest from infinity), we have
\be
\tau(r_0)\;=\;{1\over 3} \left ( (r_0+M)\sqrt{2r_0-M\over M}-2M\right ) .
\label{qtwo}
\ee
The relevance of this well-known fact for our purposes is as follows. Consider a string worldsheet in the throat of the geometry at some position $r_0>M$. Consider a fluctuation in which a spike with circumference $c$ emerges from the worldsheet. If this spike has a proper length $\tau$ larger than (\ref{qtwo}), then it can enter the region $r<M$ inside the horizon. The action for such a spike is 
\be
\Delta S \;\sim\; TA \;\sim\; {1\over \alpha'} c \:\! \tau
\ee
where $T={1\over 2\pi\alpha'}$ is the string tension, $A$ is the area of the spike and where we have used the Nambu-Goto expression to estimate the action. Since $\Delta S$ is finite, we see that the region exterior to the horizon, $r>M$, {\it does not form a  complete string background by itself}.  Even though the throat is infinite along a constant $t$ hypersurface, fluctuations of the worldsheet will explore the interior of the horizon if we assume that the geometry is that of the standard black hole.

\subsection{Touching the singularity}

Although the exterior of the horizon is not a complete string background, it might appear that this is not a serious problem: one could just consider the continuation of the metric (\ref{qone}) to the region $r<M$. But as we will now see, a finite-action spike from a worldsheet in the throat can reach the singularity. This would indeed signal a difficulty, since we cannot get a complete string background without understanding what happens at the singularity. 

It is well known that the singularity of a Reissner-Nordstrom black hole is repulsive: massive particles falling freely towards $r=0$ turn back before reaching $r=0$ and enter a new region of the extended spacetime. However when we are considering the quantum theory, we must consider all paths that have a finite action. We will now show that finite-action paths exist that connect a point at $r=r_0 >0$ to the singularity at $r=0$.

Note that we can use the coordinates in (\ref{qone}) in the region $r<M$ with the time coordinate now being denoted $t'$ to emphasize that it is not a direct continuation of $t$ outside the hole:
\be
ds^2\;=\;-f(r) \:\! dt'^2+{dr^2\over f(r)}+r^2\:\! d\Omega_2^2 \,, \qquad\quad f(r) \;=\; \left(1-{M\over r}\right)^{\! 2} \,.
\label{qthree}
\ee
Let us make a finite-action path that goes from a radius $\bar r<M$ to the singularity $r=0$ along a radial direction. The path length is
\be
\tau\;=\;\int\limits_{r=0}^{\bar r} \left (f(r) \:\! dt'^2-{dr^2\over f(r)} \right )^{\! \h}\,=\,\int\limits_{r=0}^{\bar r} dr \left (f(r) \left({dt'\over dr}\right)^{\! 2}-{1\over f(r)} \right )^{\! \h} \,.
\ee
The function $f(r)$ diverges at $r=0$ and, assuming that ${dt'\over dr}$ remains finite, this is the only potential source of divergence in this integral. We may thus take $\bar r\ll M$ for convenience, so that $f(r) \approx {M^2\over r^2}$. For concreteness let us make the choice ${dt'\over dr}=\left({r\over M}\right)^{\!\!\; \h}$, which gives
\be
\tau \;\simeq\;  \int\limits_{r=0}^{\bar r} dr\left ( {M^2\over r^2}\left({dt'\over dr}\right)^2-{r^2\over M^2}\right )^{\! \h}  \;\simeq\; \int\limits_{r=0}^{\bar r} dr {M\over r} {dt'\over dr} 
 \;=\; \int\limits_{r=0}^{\bar r} dr \left({M\over r}\right)^{\!\h} \,.
\ee
As mentioned above,  this gives a finite result,
\be
\tau ~\approx~ 2M^\h \bar r^\h \,.
\label{qfour}
\ee

In string theory, we can again consider a spike on a worldsheet that has a circumference $c$ and a length $\tau$. From the finiteness of (\ref{qfour}) we see that a finite-action spike from the worldsheet can connect a point at $0<r<M$ to $r=0$. We have already seen that a finite-action tube can connect the throat to the region $r<M$. 

Thus we conclude that finite-action spikes connect the worldsheet in the exterior near-horizon throat to the singularity, and if we do not have a boundary condition at this singularity then we cannot define a complete string background.

\subsection{Boundary conditions at the singularity and fuzzballs}

Let us now see what boundary conditions one can have for the string worldsheet at a singularity. The simplest example is given by a D-brane sitting in the spacetime. The string does not see the D-brane at leading order in the string coupling $g_s$. However if we are to describe a microstate like a fuzzball, then the solution has to exist in the full  string theory, and  therefore we need to ask how the D-brane will be 
handled in the description of the string background. In the case of the singularity describing a D-brane source,  the answer is known: the string worldsheet can have a hole with the boundary of the hole ending on the D-brane; each such hole gives a factor of the string coupling $g_s$ in the amplitude.

Thus the singularity arising from a stack of parallel D-branes is an allowed singularity corresponding to a physical source in string theory. But a single kind of charge does not make a black hole. To get a black-hole-like entropy that grows with the number of branes in the bound state, we need at least two charges; for example we can take \donep. 

Suppose we take a spherically symmetric two-charge BPS black hole and ask that the singularity at $r=0$ carry both D1 and P charges. Then there is no consistent boundary condition at this singularity. The momentum on a D-string must be carried by transverse modes, not by a longitudinal mode. Once we consider the transverse vibrations on the D1-brane, the geometry becomes precisely that of a fuzzball solution in the \donep frame. In such a fuzzball solution we find that there is no horizon and the only singularity is that of the D1-brane source. As explained in Section \ref{sec:local-dipole-charges}, at any point along the D1-brane, we see just a straight segment of the D1, moving in a direction perpendicular to itself. After going to a local frame where  the boost is removed, the singularity becomes that of a single type of charge -- the D1.\footnote{The source singularity in solutions like the extremal D1-P solutions is also a locus of infinite redshift; i.e. $g_{tt}=0$ at this source. This ensures that  virtual strings emitted from this source do not produce a gravitational field in addition to the metric  already describing in the solution. More precisely, the amplitude to emit a virtual open or closed string from the D1 source involves an integral over the worldvolume of the D1, and the vanishing of $g_{tt}$ makes this integral vanish. For the traditional black hole, on the other hand, we do not have any way to describe the singularity in terms of emitted virtual strings (since this is not a known source singularity of string theory), and so we cannot proceed to ask what effect this singularity may have on objects approaching it.}

A similar behavior occurs for more complicated fuzzballs that have three or four conserved charges. There is no consistent boundary condition in string theory for an object carrying multiple charges, but in the fuzzball the charges are resolved into different constituents which each carry a single allowed physical charge of the theory. These allowed charges are vectors in charge space which are partly along the overall conserved charges of the bound state and partly along the local dipole charges. In the overall fuzzball microstate the local dipole charges cancel, while the conserved charges add up to give the overall net conserved charges of the microstate. Importantly, the resolution of the total charge into allowed charges breaks spherical symmetry and results in a structure whose size spreads out in such a way that there is no horizon for any of the states that have been studied.

To summarize, the supersymmetric two-charge black hole solution with horizon used in the argument of \cite{Sen:2009bm} is not a complete and consistent string theory background due to the singularity, and resolving the singularity leads us to the known fuzzball microstates that have no horizon.

\subsection{A general argument for absence of horizons in extremal states}

\begin{figure}[t!]
\begin{center}
\includegraphics[width=16cm]{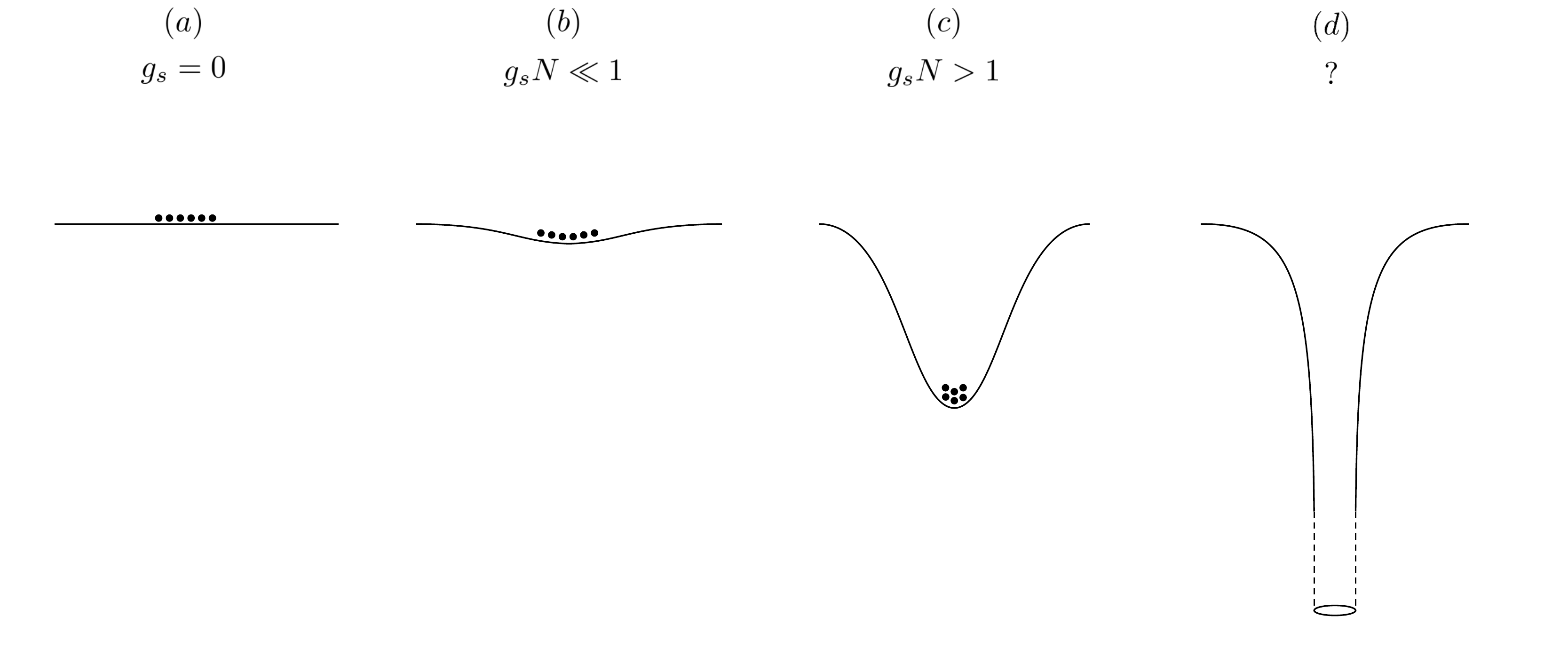}
\caption{(a)--(c) Backreaction of brane bound states at different values of $g_s$.  (d) Traditional solution with infinite throat and horizon.}
\label{fone}
\end{center}
\end{figure}

In string theory, one can make a general argument that extremal black hole solutions with throats of infinite proper length
(and horizons and the end of this throat) are not the correct description of the brane bound states that are usually assumed to correspond to such solutions. To do so, we shall employ an argument similar to the one described in Section \ref{secpt}. The string coupling $g_s$ is a continuous modulus that we are free to dial. Consider a bound state of strings and branes and let the number of these constituent objects be denoted schematically by $N$. Consider first the limit $g_s \r 0$, in which we ignore backreaction so that we simply have a finite-energy configuration in flat spacetime, as depicted in Fig.\:\ref{fone}(a). 

Now let us increase $g_s$ to a small value, and denote by $g_s N^\alpha$ the combination that controls the backreaction.
Here $\alpha$ is a positive number that depends on the particular bound state under consideration.\footnote{For instance, for a two-charge system $N^\alpha$ may stand for the combination $\sqrt{n_1n_2}$. See e.g.~\cite{Giusto:2004xm} Eq.\;(2.39).}
When $g_s N^\alpha \ll 1$, the backreaction of the bound state on the spacetime can be computed and is non-zero but small everywhere, as depicted in Fig.\:\ref{fone}(b). 

As we dial $g_s$ continuously to larger values, the backreaction increases continuously. Let us next consider $g_s N^\alpha $ to be bigger than 1, but not several orders of magnitude bigger. Then the backreaction becomes order one or larger in a `throat' region surrounding the branes, as depicted in Fig.\:\ref{fone}(c). (For example in the two-charge system this can be read off from the solutions (\ref{eq:ns1p-sol}), (\ref{D1D5Chiral}), (\ref{D1Psol}).)
The proper length $L(g_s N^\alpha)$ of this throat is expected to be a continuous function of $g_sN^\alpha$.

When $g_s N^\alpha$ is several orders of magnitude bigger than 1, such that the backreaction is order one or larger over macroscopic lengthscales, we are in the domain where we expect this configuration to correspond to a macroscopic extremal black hole.
However, also in this range of values of $g_s N^\alpha$, and indeed over the full range of non-zero values of $g_s N^\alpha$ up to arbitrarily large values, the proper length of the throat $L(g_s N^\alpha)$ is expected to be continuous. Note that arbitrarily large $g_s N^\alpha$ can be achieved while dialling $g_s$ only within the perturbative regime $g_s \ll 1$ provided of course that one first fixes $N$ to be appropriately large. 
In particular, this would imply that the length of the throat cannot diverge at any finite $g_s N^\alpha$. 
By this reasoning one expects that the traditional geometry depicted in Fig.\;\ref{fone}(d), where the proper length of the throat is infinite, is never the correct description of the bound state.

In order to show the opposite, that the length of the throat becomes infinite, one would have to demonstrate a phase transition for the states of the extremal black hole at some finite critical value of $g_s N^\alpha$, such that this phase transition results in a divergence of $L(g_s N^\alpha)$ at the critical value. We are not aware of any instance where such a divergence has been demonstrated. 

The situation depicted in Fig.\:\ref{fone}(c) represents the fuzzball paradigm for extremal holes also in the regime where $g_s N^\alpha$ is several orders of magnitude bigger than 1: there is a long but finite throat, and the detailed physics of the `cap' region deep inside in this geometry contains the information of the bound state, possibly involving highly stringy and/or quantum physics. In particular, the state is not well described by a geometry with horizon, and thus there is no region interior to the horizon.

\section{The significance of the solution with horizon}
\label{sec:trad-sol-interpretation}

We have argued above the microstates of the black hole should be fuzzballs, with no horizon. What then should we make of the traditional black hole solution that we obtain from classical physics, which does have such a horizon?

First, we note that this solution with horizon was obtained by imposing the assumption of spherical symmetry, and then solving the resulting field equations. We have seen that at least for the two-charge supersymmetric bound states, the actual microstates break the symmetries of the traditional black hole solution. In principle there would be nothing wrong with assuming a spherically symmetric ansatz if this ansatz resulted in  a regular solution, or a solution that is regular apart from allowed sources in string theory; in this situation we would have to accept the solution as a valid one. However we have seen that the solution we obtain this way has a singularity that is not allowed in string theory, and so it is not a valid string background. 

Can one still understand the traditional solution with horizon in some way? Consider the (3+1)-dimensional Reissner-Nordstrom hole with mass $M$ and charge $Q$. This has a temperature
\be
T\;=\;{1\over 2\pi}{\sqrt{M^2-Q^2}\over  (M+\sqrt{M^2-Q^2})^2} \,.
\ee
Euclideanizing time $t\r -i\tau$, we obtain the Euclidean Reissner-Nordstr\"{o}m solution,
\be
ds^2\;=\;\left(1-{2M\over r}+{Q^2\over r^2}\right) d\tau^2 + {dr^2\over \left(1-{2M\over r}+{Q^2\over r^2}\right)}+r^2d\Omega_2^2 \,.
\label{qten}
\ee
If we compactify $\tau$ as $\tau\sim \tau+{1\over T}$, then one obtains a solution that caps off smoothly at  
the outer horizon, $r=r_+=M+\sqrt{M^2-Q^2}$. We can thus consider the path integral of the gravity theory over all solutions with mass $M$ and charge $Q$, in a loop with Euclidean time period ${1\over T}$. We expect $\,\exp[S_{\mathrm{bek}}]$ microstates, each contributing $\,\exp[-{M\over T}]$  to the loop integral. This path integral can be approximated by $\,\exp[-I]$, where $I$ is the action of the solution (\ref{qten}) computed with the standard Gibbons-Hawking-York boundary term at infinity. 

This is the method that was used by Gibbons and Hawking to compute black hole entropy~\cite{Gibbons:1976ue}. A smooth Euclidean solution such as (\ref{qten}) is a gravitational instanton, which gives a saddle point about which one can expand, to compute the on-shell action and thereby the entropy. Sen follows a related procedure in \cite{Sen:2007qy}, 
focusing on the extremal limit $Q\r M$. (In this process $T\r 0$, and the limit needs to be taken with some care.) The Wald entropy used by Sen is an expansion to arbitrary orders around the saddle point.

We thus obtain a satisfactory interpretation of the traditional spherically symmetric geometry. The actual microstates (in Lorentzian spacetime) are fuzzballs without horizon. One may count these microstates using a one-loop path integral over Euclidean time. This path integral may be approximated by a Euclidean saddle point solution, and successive corrections to the count of states will be given by expanding around this saddle point. C.f.~the remarks on this point in~\cite{Skenderis:2008qn,Mathur:2014nja}. Since the saddle point solution represents all the fuzzball states, it should be expected to be a spherically symmetric Euclidean solution. 

One may continue the Euclidean saddle point solution to Lorentzian signature. 
However we re-emphasize that the resulting Lorentzian solution with horizon has an unphysical singularity inside the horizon which is not a valid source in string theory. In addition, non-extremal black hole solutions with horizons give rise to the Information Paradox~\cite{Hawking:1976ra,Mathur:2009hf}. The formal computation of entropy appears to be the same whether we use the Lorentzian solution or the Euclidean one. This leads to the apparent direct physical meaning of the Lorentzian solution with horizon, but we have argued that this solution should really be understood through its Euclidean continuation. 

Several computations have been carried out using the extended spacetime of the eternal black hole in AdS space, whereby correlation functions between operators on the two different boundary CFTs have been reproduced by dual computations in the extended spacetime. However the correlators compared here were those that can be obtained by analytic continuation from Euclidean to Lorentzian; see \cite{Kraus:2002iv} for examples of such continuations. Thus the agreement cannot be taken as a proof of existence of the Lorentzian black hole with horizons: if we have an agreement between a Euclidean CFT correlator and a computation in a Euclidean dual geometry, and then we analytically continue both sides of the duality to Lorentzian time, the agreement will naturally continue to hold.\footnote{The traditional black hole solution should not, in our opinion, be thought of as arising from averaging over fuzzball solutions to obtain an ``average geometry''. There is no clear way to average two metrics, since there is no canonical way to associate points in two different geometries; further, there is no canonical choice of coordinates in which the metrics should be written before averaging.  The one-loop Euclidean path integral described above is the natural way in which one can see all the fuzzball solutions at the same time.}  However it has been argued that other comparisons between the boundary CFTs and the extended eternal black hole will not work, and that the extended eternal black hole should not exist as a stable solution in the gravitational theory~\cite{Mathur:2014dia}.

In~\cite{Sen:2009bm} Sen also argues that the wavefunction of a string bound state can change character as one moves around in moduli space.  While this is true for any quantum theory, here we simply note that unitarity will not allow the many different wavefunctions corresponding to fuzzball states to change into to just one wavefunction describing the traditional hole with horizon. 

It is natural to expect that typical fuzzball microstates will have their nontrivial structure inside the classical horizon radius $r_h$ and perhaps one or a few Planck lengths outside, but for $r>>r_h+l_p$ we expect the fuzzball solution to rapidly approach the corresponding traditional black hole solution. Such a picture of was described as a ``tight'' fuzzball in \cite{Guo:2017jmi}, and an argument was given that generic fuzzballs should indeed be tight in this sense.  With this picture, we see that observations made away from the horizon will lead to the same behavior as expected from the classical black hole. At the same time the large quantum deviations from the classical solution inside the radius $r_h$ resolve the information paradox. 

As has been noted in the firewall argument \cite{Almheiri:2012rt}, there will be high-energy radiation very close to the fuzzball, in contrast to the traditional black hole which has the (say Unruh) vacuum around the horizon. In particular at Planck distances from the horizon, the temperature of this radiation will be Planck-scale, and the associated gravitons will of course give strong quantum fluctuations in the metric. As has been emphasized several times in other articles, this strongly quantum nature of the generic fuzzball is expected to continue into the fuzzball interior as well (see e.g.~\cite{Mathur:2012jk,Mathur:2013gua,Guo:2017jmi}). However as noted in Section \ref{sec:limit-to-generic}, the important point is that this generic quantum solution is obtained as a limit of solutions where the evolution of outgoing modes of Hawking radiation can be studied, and the evolution is seen to differ by order unity from the evolution in the vacuum solution. This resolves the information paradox in the fuzzball paradigm. By contrast, in certain alternative proposed resolutions, the horizon is smooth, but speculative nonlocal effects extending far outside the horizon are introduced; see e.g.~\cite{Giddings:2012gc,Papadodimas:2012aq,Maldacena:2013xja}.

\section{Discussion}
\label{sec:disc}

In this paper we have argued that the black hole with horizon used by Sen in his computation of entropy is best viewed as a technical tool to compute the black hole entropy, and that the solution is more correctly interpreted as a Euclidean saddle-point solution rather than a solution describing a microstate of the black hole. In particular we have argued that the microstates in all duality frames are fuzzballs, in contrast to the perspective in \cite{Sen:2009bm} which has been interpreted by some as implying that the microstates are fuzzballs in the \donedfive frame but a black hole with horizon in the \nsonep frame. 
Our argument was twofold. First, we observed that the logic used in \cite{Sen:2009bm} to distinguish the \donedfive and \nsonep frames fails when applied to the \donep and other duality frames. Second, the black hole solution does not provide a valid string solution because of the unphysical singularity inside the horizon, and resolving the singularity for two-charge bound states brings the solution back to a fuzzball in every duality frame. 

The fuzzball proposal provides a consistent resolution of the information paradox. To understand the proposal however, one must appreciate the nature of the difficulties we face in seeking a complete quantum description of black holes. 

First, one must fully appreciate the `no-hair' nature of the horizon. 
While many no-hair theorems are classical, results such as those of Price \cite{Price:1971fb,*Price:1972pw} show that under very general assumptions, the state of a quantum field around the horizon sheds its information and tends to the (say Unruh) vacuum state on the timescale of the light-crossing time of the black hole. Remarkably, the fuzzball constructions in string theory describe structures that support themselves against gravitational collapse even we we add non-extremality~(see e.g.~\cite{Jejjala:2005yu,Mathur:2013nja,Bossard:2014ola,Bena:2015drs,Bena:2016dbw,Bossard:2017vii}); this behavior may be traced to a change in the topology whereby the compact directions become fibered nontrivially over the noncompact ones~\cite{Gibbons:2013tqa,deLange:2015gca}. A toy model explaining this phenomenon was given in \cite{Mathur:2016ffb}.

Second, we expect that all microstates of the black hole are fuzzballs, and we expect that the generic fuzzball will have structure at the Planck scale. This is the case for the two-charge fuzzballs, where we can understand all the states in terms of the vibrations of a string. The expected Planck-scale structure of microstates provides a technical barrier to understanding typical states explicitly in a quantum theory of gravity. However in string theory we have found a way to order the microstates in terms of their complexity, as explained in Section \ref{sec:limit-to-generic}. This allows a limiting process in which one can see that non-generic fuzzballs do not have a horizon with the vacuum around this horizon, and importantly, that the limit towards generic states takes us {\it away} from the vacuum rather than towards it, since the fuzzball structure becomes more complicated the more generic the state becomes. It is sometimes stated that since we cannot understand generic states in a reliable supergravity description, it is unclear what we can learn from fuzzballs (see e.g.~\cite{Chen:2014loa,Raju:2018xue}).  This is a misconception: what we learn is how the no hair results can be bypassed in string theory to yield states where the would-be interior of the black hole is not a vacuum region, and there is no region that is causally disconnected from future infinity. If one would like to show that the fuzzball construction is not a resolution of the information paradox, then one would have to demonstrate a phase transition in the space of microstate solutions where the simple non-generic states change over to a qualitatively different structure with a vacuum horizon, and where the wavefunction of the bound state that resolves the singularity is confined well within this horizon, as we approach generic states. However we are not aware of any such computation.

Third, the process that gives rise to Hawking radiation from a fuzzball is not well-approximated by the black hole with horizon. More precisely, the evolution of the Hawking modes will deviate by {\it order unity} from the evolution of the corresponding modes in vacuum. (For two-charge fuzzballs this can be seen by looking at the scalar wave-equation in the fuzzball background, or by studying classical geodesics in the fuzzball 
\cite{Lunin:2002qf} and using their paths in an eikonal approximation.) Models in which the process of Hawking radiation does not have such a deviation but that attempt to restore unitarity tend to introduce speculative new physics, such as non-local effects that modify the Hawking radiation in the near-horizon region~\cite{Giddings:2012gc} or when it has escaped far from the black hole~\cite{Papadodimas:2012aq,Maldacena:2013xja}. It remains highly unclear whether such effects can be derived from string theory.

Finally, let us comment on the `infall problem': what does an infalling observer feel as they fall onto the fuzzball? One might naively imagine that the observer will see the detailed structure of the fuzzball interior, and therefore cannot effectively experience free fall through a vacuum region; however this would be too simplistic. Let us assume that the infalling observer has energy $E\gg T$ at infinity. In this situation the added energy creates {\it new} fuzzball states corresponding to an added entropy
\be
\Delta S \;\sim\; {E\over T} \; \gg \; 1
\ee
so that there is a much larger number $\,\exp[S_{\mathrm{bek}}+\Delta S]\,$ of new fuzzball states
than the number $\,\exp[S_{\mathrm{bek}}]\,$  of fuzzball states of the original black hole. The conjecture of fuzzball complementarity states that the evolution of the quantum gravity wavefunctional in this superspace of new fuzzball solutions can be mapped to an approximately free infall for the infalling observer~\cite{Mathur:2012jk,Mathur:2013gua}; for a recent discussion, see~\cite{Mathur:2017fnw}. In particular, in order to investigate the infall of an $E\gg T$ quantum, one must take account of the fact that the infall will cause the original fuzzball solution to transition into a band of newly available states; a model was presented in \cite{Mathur:2015nra} for how evolution in this band can mimic free infall.

\vspace{2mm}

\section*{Acknowledgements}

We thank Stefano Giusto, Keshav DasGupta, Emil Martinec, Rodolfo Russo, Masaki Shigemori, Kostas Skenderis, Marika Taylor and especially Ashoke Sen for discussions and/or comments on earlier versions of the manuscript. We also thank the authors of~\cite{Cano:2018hut} for communicating a draft of their work prior to publication. The work of SDM is supported in part by the DOE grant DE-SC0011726. The work of DT is supported by a Royal Society Tata University Research Fellowship. DT acknowledges the hospitality of CEA Saclay and CERN during the course of this work.


\vspace{2mm}

\begin{adjustwidth}{-2mm}{-2mm} 

\baselineskip=14pt
\parskip=2.5pt

\bibliographystyle{utphysM}      

\bibliography{microstates}       

\end{adjustwidth}


\end{document}